\documentclass{emulateapj}
\usepackage{natbib}
\usepackage{lscape}
\usepackage{multirow}
\def\CIVdblt{{\rm C~}\kern 0.1em{\sc iv}~$\lambda\lambda 1548, 1550$}
\def\MgIIdblt{{\rm Mg~}\kern 0.1em{\sc ii}~$\lambda\lambda 2796, 2803$}
\def\NVdblt{{\rm N}\kern 0.1em{\sc v}~$\lambda\lambda 1238, 1242$}  
\def\OVIdblt{{\rm O}\kern 0.1em{\sc vi}~$\lambda\lambda 1031, 1037$}
\def\SiIVdblt{{\rm Si~}\kern 0.1em{\sc iv}~$\lambda\lambda1394, 1403$}
\def\AlIIIdblt{{\rm Al~}\kern 0.1em{\sc iii}~$\lambda\lambda1855,1863$}
\def\FeIIdblt{{\rm Fe~}\kern 0.1em{\sc ii}~$\lambda\lambda 2383, 2600$}
\def\NeVIIIdblt{{\rm Ne~}\kern 0.1em{\sc viii}~$\lambda\lambda 770, 780$}

\def\NeVIII{\hbox{{\rm Ne~}\kern 0.1em{\sc viii}}}
\def\OI{\hbox{{\rm O~}\kern 0.1em{\sc i}}}
\def\OII{\hbox{{\rm O~}\kern 0.1em{\sc ii}}}
\def\OIII{\hbox{{\rm O~}\kern 0.1em{\sc iii}}}
\def\OIV{\hbox{{\rm O~}\kern 0.1em{\sc iv}}}
\def\OVI{\hbox{{\rm O~}\kern 0.1em{\sc vi}}}
\def\OVII{\hbox{{\rm O~}\kern 0.1em{\sc vii}}}
\def\OVIII{\hbox{{\rm O~}\kern 0.1em{\sc viii}}}
\def\NIII{\hbox{{\rm N~}\kern 0.1em{\sc iii}}}
\def\NIV{\hbox{{\rm N~}\kern 0.1em{\sc iv}}}
\def\NVII{\hbox{{\rm N~}\kern 0.1em{\sc vii}}}
\def\CIII{\hbox{{\rm C~}\kern 0.1em{\sc iii}}}
\def\SiIII{\hbox{{\rm Si~}\kern 0.1em{\sc iii}}}
\def\SVI{\hbox{{\rm S~}\kern 0.1em{\sc vi}}}
\def\NeIX{\hbox{{\rm Ne~}\kern 0.1em{\sc ix}}}

\def\AlII{\hbox{{\rm Al~}\kern 0.1em{\sc ii}}}
\def\AlIII{\hbox{{\rm Al~}\kern 0.1em{\sc iii}}}
\def\CaI{\hbox{{\rm Ca}\kern 0.1em{\sc i}}}
\def\CaII{\hbox{{\rm Ca}\kern 0.1em{\sc ii}}}
\def\CrII{\hbox{{\rm Cr}\kern 0.1em{\sc ii}}}
\def\CII{\hbox{{\rm C~}\kern 0.1em{\sc ii}}}
\def\CIII{\hbox{{\rm C~}\kern 0.1em{\sc iii}}}
\def\CIV{\hbox{{\rm C~}\kern 0.1em{\sc iv}}}
\def\CV{\hbox{{\rm C}\kern 0.1em{\sc v}}}
\def\H{\hbox{{\rm H}}}
\def\HI{\hbox{{\rm H~}\kern 0.1em{\sc i}}}
\def\HII{\hbox{{\rm H~}\kern 0.1em{\sc ii}}}
\def\Lya{\hbox{{\rm Ly}\kern 0.1em$\alpha$}}
\def\Lyb{\hbox{{\rm Ly}\kern 0.1em$\beta$}}
\def\Lyg{\hbox{{\rm Ly}\kern 0.1em$\gamma$}}
\def\Lyth{\hbox{{\rm Ly}\kern 0.1em$\theta$}}
\def\Lyfive{\hbox{{\rm Ly}\kern 0.1em$5$}}
\def\Lysix{\hbox{{\rm Ly}\kern 0.1em$6$}}
\def\Lyseven{\hbox{{\rm Ly}\kern 0.1em$7$}}
\def\Lyeight{\hbox{{\rm Ly}\kern 0.1em$8$}}
\def\Lynine{\hbox{{\rm Ly}\kern 0.1em$9$}}
\def\Lyten{\hbox{{\rm Ly}\kern 0.1em$10$}}
\def\HeI{\hbox{{\rm He}\kern 0.1em{\sc i}}}
\def\HeII{\hbox{{\rm He}\kern 0.1em{\sc ii}}}
\def\FeI{\hbox{{\rm Fe~}\kern 0.1em{\sc i}}}
\def\FeII{\hbox{{\rm Fe~}\kern 0.1em{\sc ii}}}
\def\FeIII{\hbox{{\rm Fe~}\kern 0.1em{\sc iii}}}
\def\MnII{\hbox{{\rm Mn}\kern 0.1em{\sc ii}}}
\def\MgI{\hbox{{\rm Mg~}\kern 0.1em{\sc i}}}
\def\MgII{\hbox{{\rm Mg~}\kern 0.1em{\sc ii}}}
\def\MgIII{\hbox{{\rm Mg~}\kern 0.1em{\sc iii}}}
\def\MgIV{\hbox{{\rm Mg~}\kern 0.1em{\sc iv}}}
\def\NaI{\hbox{{\rm Na}\kern 0.1em{\sc i}}}
\def\NV{\hbox{{\rm N}\kern 0.1em{\sc v}}}
\def\NII{\hbox{{\rm N}\kern 0.1em{\sc ii}}}
\def\NIII{\hbox{{\rm N}\kern 0.1em{\sc iii}}}
\def\OVI{\hbox{{\rm O}\kern 0.1em{\sc vi}}}
\def\SiII{\hbox{{\rm Si~}\kern 0.1em{\sc ii}}}
\def\SiIII{\hbox{{\rm Si~}\kern 0.1em{\sc iii}}}
\def\SiIV{\hbox{{\rm Si~}\kern 0.1em{\sc iv}}}
\def\SII{\hbox{{\rm S}\kern 0.1em{\sc ii}}}
\def\SIII{\hbox{{\rm S}\kern 0.1em{\sc iii}}}
\def\SIV{\hbox{{\rm S}\kern 0.1em{\sc iv}}}
\def\TiII{\hbox{{\rm Ti}\kern 0.1em{\sc ii}}}
\def\ZnII{\hbox{{\rm Zn}\kern 0.1em{\sc ii}}}
\newcommand{\kms}{\hbox{km~s$^{-1}$}}
\newcommand{\cmsq}{\hbox{cm$^{-2}$}}
\newcommand{\cc}{\hbox{cm$^{-3}$}}
\def\kms{\hbox{km~s$^{-1}$}}      
\def\cmsq{\hbox{cm$^{-2}$}}
\def\cc{\hbox{cm$^{-3}$}}
\def\etal{et~al.\ }

\begin{document}

\title{Detection of {\NeVIII} in an Intervening Multi-Phase \\ Absorption System Towards 3C~263\altaffilmark{1}}
\author{Anand Narayanan\altaffilmark{2}, Bart P. Wakker\altaffilmark{2}, Blair D. Savage\altaffilmark{2}}

\altaffiltext{1}{Based on observations with the NASA-CNES-CSA {\it Far Ultraviolet Spectroscopic Explorer} operated by Johns Hopkins University, supported by NASA contract NAS5-32985.}
\altaffiltext{2}{Department of Astronomy, The University of Wisconsin-Madison Email: anand, wakker, savage@astro.wisc.edu}

\begin{abstract}
 
We report on the detection of {\NeVIII} in an intervening multiphase absorption line system at $z=0.32566$ in the FUSE spectrum of the quasar 3C~263 ($z_{em} = 0.646$). The {\NeVIII}~$\lambda 770$~{\AA} detection has a 3.9~$\sigma$ significance. At the same velocity we also find absorption lines from {\CIV}, {\OIII}, {\OIV} and {\NIV}. The line parameter measurements yield log~[$N({\NeVIII})~{\cmsq}] =13.98^{+0.10}_{-0.13}$ and $b = 49.8~\pm~5.5$~{\kms}. We find that the ionization mechanism in the gas phase giving rise to the {\NeVIII} absorption is inconsistent with photoionization. The absorber has a multi-phase structure, with the intermediate ions produced in cool photoionized gas and the {\NeVIII} most likely in a warm collisionally ionized medium in the temperature range $(0.5 - 1.0) \times 10^6$~K. This is the second ever detection of an intervening {\NeVIII} absorption system. Its properties resemble the previous {\NeVIII} absorber reported by Savage {\etal}(2005). Direct observations of {\HI} and {\OVI} are needed to better constrain the physical conditions in the collisionally ionized gas phase of this absorber. 
 
\end{abstract}


\section{INTRODUCTION}

Cosmological hydrodynamical simulations of the gravitational assembly of matter suggest that most of the baryons in our universe exist in the intergalactic medium in multiple gas phases of temperatures and densities, with the mass fraction in each phase evolving with redshift \citep{cen99, dave99, dave01, cen06}. Among these phases the shock-heated warm-hot intergalactic medium (WHIM) is highly significant, since it bears approximately 30~\% -- 50~\% of the baryonic mass fraction at low-$z$ \citep{dave99, cen06}. The WHIM gas is predicted to exist in the {\it warm} ($10^5 - 10^6$~K) and {\it hot} ($10^6 - 10^7$~K) temperature ranges \citep{cen99, dave99}. Even though it is expected to be the dominant reservoir of baryons at low-$z$, WHIM detections have been limited in the past. The shock heated gas is likely heavily ionized ($f_{\HI} =~$N(\HI)/N(\H)$~\sim 10^{-6}$) and hence spectroscopic observations of high ionization species or very broad Ly-$\alpha$ lines are required to trace this gas phase in the intergalactic medium (IGM). 

In the UV wavelength regime, intergalactic {\OVI} has been used as a probe of the warm component of the WHIM. For an assumed [O/H] $\sim -1$ elemental abundance in the absorbing gas, it has been estimated that the {\OVI} intergalactic absorbers potentially account for 5~\% - 10~\% of the baryon budget in the low-$z$ universe \citep{danforth05}. However, not all {\OVI} absorbers can be treated as tracers of the WHIM phase of the IGM. The ion can be produced in both cool photoionized gas as well as in a warm collisionally ionized medium \citep{savage02, danforth05}. Due to this, it has proven difficult to establish the actual temperature in the complex {\OVI} absorbers that could be associated with warm WHIM gas \citep{lehner06, tripp08}. 
 
Absorption at X-ray energies by highly ionized atoms such as {\OVII}, {\OVIII}, {\NVII}, and {\NeIX} are best suited for probing the hot phase of the WHIM. Such gas phases are expected to be part of large scale filaments connecting to the over-dense galaxy cluster and group environments. Compared to the detections of the warm phase of the WHIM in the UV and FUV, the X-ray observations of the highly ionized WHIM at $z > 0$ have been less successful primarily due to instrumental limitations (insufficient spectral resolution and sensitivity). A few detection claims for gas with $z > 0$ were reported in the past \citep{fang02, nicastroa, nicastrob}, although the validity of those detections has been challenged \citep{kaastra06, bregman07, richter08}.

The strong resonance transitions of {\NeVIIIdblt}\footnote{The wavelengths throughout are given as vacuum wavelengths rounded to the nearest natural number.} are potentially secure probes of collisionally ionized gas at T $\sim (0.5 - 1.0) \times 10^6$~K \citep{savage05}. However, few firm detections of {\NeVIII} exists, primarily due to the observational limitations in the UV and FUV. \citet{savage05} reported the first~$\geq 3$~$\sigma$ detection of {\NeVIII} in an intervening absorber. The lines were detected at $z = 0.20701$, along the sight line to the quasar HE~$0226-4110$. The combined FUSE and STIS spectrum of this target facilitated observations of a host of low, intermediate and high ionization metal lines and associated {\HI} allowing robust constraints for the multiple gaseous phases in the absorber. \citet{prochaska04} listed a 3.7~$\sigma$ significance {\NeVIII} detection in the $z = 0.49510$ metal line system towards PKS~$0405-123$. With improved $S/N$ FUSE data for this sight line, we find that the {\NeVIII} is in fact a non-detection at the 3~$\sigma$ level, in agreement with the more complete analysis of this absorber by \citet{howk08}. We measure a rest-frame equivalent width of $W_r({\NeVIII}~\lambda 770) = 10.1 \pm 8.9$~m{\AA}, from integrating over the velocity window corresponding to the {\OVI} for this absorber. Other than the two cases above, only upper limits for {\NeVIII} based on 3~$\sigma$ non-detections have been reported for metal line absorption systems \citep[e.g.][]{richter04,lehner06}.

The analysis presented in \citet{savage05} showed that detectable amounts of {\NeVIII} are likely to be created in gas only under collisional ionization conditions with $T \sim (0.5 - 1.0) \times 10^6$~K. In the $z=0.20701$ absorber, they found the origin of the intermediate ionization species such as {\CIII}, {\NIII}, {\OIII}, {\SiIII}, {\OIV} and {\SVI} to be consistent with a photoionized medium of low total hydrogen density with $n_{\H} = 2.6 \times 10^{-5}$~{\cc} and $N(\H) = 4.6 \times 10^{18}$~{\cmsq}, and relatively modest neutral fractions ($f_{\HI} = 2.5 \times 10^{-4}$). However, {\OVI} and {\NeVIII} required the presence of gas in an entirely different phase that was collisonally ionized, with a temperature of T$ = 5.4 \times 10^5$~K, indicating that these ions were tracing the warm shock heated gas. Unlike {\OVI} which can arise in varied ionization conditions \citep[e.g.][]{tripp01,prochaska04,danforth05,tripp08,oppenheimer08}, the presence of {\NeVIII} entails warm gas that is collisionally ionized \citep{savage05}.  The ion is hence a more reliable probe for detecting warm collisionally ionized gas in the low-$z$ universe. 

Identifying the environments where absorbers reside is important for understanding their origin.  Obtaining the redshift of galaxies in the general direction of QSOs whose spectra contain interesting absorption systems can provide information about what lies in the vicinity of those absorbers.   The absorber/galaxy association  studies of \citet{stocke06} and \citet{wakker09} reveal that {\OVI} absorbers are clearly associated with the extended environments of galaxies.  The sites traced by these absorbers can be the extended halos of galaxies, intra-group gas, or WHIM filaments containing metals that connect to the environments of galaxies.  Discriminating among these different possible absorbing sites requires careful work.  In the case of the  {\NeVIII} system at $z = 0.20701$ seen toward HE~$0226-4110$ \citep{savage02},  the recent redshift study and analysis effort of \citet{mulchaey09} has shown that the {\NeVIII} absorption system probably occurs  in a cool-hot gas interface in  the hot  extended halo of an 0.25$L^*$ galaxy situated at an impact parameter of 77 kpc from the sight line.  Therefore,  detecting absorbers with physical conditions consistent with shock heated collisionally ionized gas does not necessarily signal the direct  detection of the WHIM.  The WHIM is heated though the release of gravitational potential energy as structures form in the universe through the gravitational assembly of matter.  The heating process continues as individual galaxies form and  can lead to the production of  highly extended hot halos around galaxies. Although the halos are not part of the WHIM,  they may  contain very significant  reservoirs of baryons and thefore are important for studying the baryonic content of the universe. 

Here we report on the detection of a {\NeVIII}~$\lambda 770$~{\AA} absorption feature in a system at $z = 0.32566$ in the {\it Far Ultraviolet Spectroscopic Explorer} (FUSE) spectrum of the quasar 3C263, making it the second ever detection of this ion at $\geq 3$~$\sigma$ significance. In Sec 2 we provide details on the reduction, analysis and wavelength calibration of the FUSE data. In Sec 3 we discuss the observed properties of the lines associated with this system. We then comment on the multiphase nature of the absorber and the dominant ionization mechanisms in it. Sec 5 is a comparison between this system and the {\NeVIII} system analyzed in \citet{savage05}. Based on this comparison, we briefly discuss the likely physical conditions in the gas traced by {\NeVIII}. 

\section{FUSE Observations}

The 3C~$263$ FUSE observations were obtained through programs E848 (7.5 ks, Sembach), D808 (3.4 ks, Sembach), G044 (53.8 ks, Shull), and F005 (196.2 ks, Savage). The spectra were processed using the CALFUSE (ver 2.4) pipeline software. The data reduction procedures are described in \citet{wakker03} and \citet{savage06}. The combined spectrum extends from 912 - 1185~{\AA}, but has low S/N for $\lambda < 1000$~{\AA}. The spectral resolution is $\sim 20$~{\kms} (FWHM). An offset in velocity for the exposures in each detector segment was derived by aligning low ionization UV ISM absorption lines to the 21-cm emission components seen in the direction of 3C~263 in the LAB survey \citep{kalberla05}. The velocity shift is required to correct for alignment of spectral features in the FUSE detector segments between the various exposures. A composite spectrum was produced by co-adding the observations at each wavelength. The final spectra for $\lambda > 1000$~{\AA} only include observations from the LiF channels, which have much higher throughput than the SiC channels at these wavelengths. For display purposes, the spectra in Figure 1 were binned to $8$~{\kms} samples corresponding to 2.5 samples per 20~{\kms} resolution element. The photon statistical S/N per 8~{\kms} sample in the 1014 - 1106~{\AA} region (corresponding to 765 - 834~{\AA} in the $z = 0.32566$ rest frame) ranges from $S/N = 5 - 10$ (see the 1~$\sigma$ photon counting error spectra in Figure 1). Global and local continua were fitted to the observations as described in \citet{wakker03}. The use of Legendre polynomials for the continua allowed for an estimate of the continuum fitting errors for each observed absorption line \citep[see][]{sembach97}. 

\section{Observed Properties of The $z = 0.32566$ Absorption Line System}

The absorption line system is detected at $z_{abs}(\OIV) = 0.32566$. It was discovered in a systematic search for all multiple-line metal absorbers in the FUSE spectrum of 3C~263. The system plot centered on the rest-frame of the absorber is shown in Figure 1. Of the {\NeVIII} doublet, the $\lambda 780$~{\AA} line is blended with Ly $\gamma$ from an absorber at $z = 0.0634$. The blend is confirmed by the detection of Ly $\beta$, Ly $\delta$, {\CIII}~$\lambda 977$~{\AA}, and {\NIII}~$\lambda 990$~{\AA} at the same redshift. On the other hand the {\NeVIII}~$\lambda 770$~{\AA} feature is distinct in the spectrum. We derive an equivalent width of $W_r({\NeVIII}~\lambda 770) =  47.0 \pm 11.9$~m{\AA} for this line in the rest-frame of the absorber. The 1~$\sigma$ uncertainty in the above rest-frame measurement incorporates the continuum placement (4.9 m{\AA}) and fixed pattern noise (5.0 m{\AA}) errors along with the statistical uncertainty (9.6 m{\AA}), and therefore is a conservative estimate of the detection significance.  The separate errors were combined in quadrature \citep[see][for details on our procedure for error analysis]{wakker03,savage05}. The fixed pattern noise estimate for the 1015 - 1025 {\AA} region of the FUSE measurement was obtained by studying the noise in some of the high S/N FUSE spectra of bright AGNs  obtained through multiple exposures with different object alignments on the detector segments.  In cases where the photon count noise was very small , the observed spectra revealed irregular noise structures at the level of $\sim 5$~m{\AA}  in the observed reference frame, which  is interpreted as fixed pattern noise. If we were to consider only the statistical uncertainty from photon counting ($9.6$~m{\AA}), then the significance of this detection would have been claimed as 4.9~$\sigma$.  The {\NeVIII}~$\lambda 770$~{\AA} feature is also independently detected in the LiF 1A and LiF 2B detector channels. In the combined LiF 1A spectrum, the line is measured to have rest-frame equivalent width of $W_r = 37.3~{\pm}~13.1$~m{\AA}, and in the lower $S/N$ LiF 2B, the line has a measured strength of $W_r = 63.2~{\pm}~21.9$~m{\AA}, over the same wavelength range. These two independent measurements are statistically compatible with each other and with the measurement derived from the combined data. 

The {\NeVIII} absorption feature is at velocity coincident with three other lines whose properties are completely consistent with an interpretation as {\OIII}~$\lambda 832$~{\AA}, {\OIV}~$\lambda 788$~{\AA}, and {\NIV}~$\lambda 765$~{\AA}. The rest-frame equivalent widths derived for these lines are listed in Table 1. The low ionization species {\OII}~$\lambda 834$~{\AA} is not detected, with a 3~$\sigma$ upper limit of 29.8 m{\AA} in the absorber rest-frame. The validity of this absorption system is further enhanced by the detection of {\CIVdblt} in the FOS G190H (FWHM = 1.5~{\AA}) spectrum at $\lambda \sim 2052.4$~{\AA}, also reported by \citet{bahcall93}. The {\CIV} doublet falls in the wings of the intrinsic {\Lya} emission, and also is adjacent to the {\NVdblt} emission feature associated with the quasar. A continuum was therefore fit locally to this absorption feature using a first order polynomial. We derive {\CIV} rest-frame equivalent widths of $W_r ({\CIV}~\lambda 1548) = 0.27~{\pm}~0.03$~{\AA}, and $W_r ({\CIV}~\lambda 1551) = 0.14~{\pm}~0.03$~{\AA}. We do not have a reliable measurement on the {\HI} associated with this absorber. The redshifted {\Lya} line at 1611.6~{\AA} is in the blue edge of the FOS G190H spectrum obtained by Bahcall {\etal}(1993) where the S/N is low. The $1~\sigma$ equivalent width uncertainty at this wavelength is 0.3~{\AA}. Based on a non-detection at the $3~\sigma$ significance level, the {\Lya} equivalent width is therefore only constrained to $W_r < 1$~{\AA}. The higher order Lyman series transitions are redshifted into the $1209-1360$~{\AA} wavelength window, while the existing FOS spectra provide coverage only for $1600-3300$~{\AA}. 

At $\lambda \sim 1015$~{\AA} ($\Delta v \sim +230$~{\kms} from the {\NIV}~$\lambda 765$~{\AA} line)  a feature  is seen with $W_{observed} = 70.0~{\pm}~13.7$~m{\AA} in the combined LiF 1A and LiF 2B measurements shown in Figure 1. This feature is not produced by the ISM and we could not identify it with any possible redshifted metal line belonging to other metal line systems in the spectrum of 3C~263.  It is not due to redshifted  Ly$\gamma$ or higher Lyman series lines, since no associated  longer wavelength Lyman lines were detected.   To verify the reliablity of the observation,  we looked separately at the  LiF 1A and LiF 2B channel measurements.   The $S/N$ ratio in the  LiF 2B channel observation is lower than for the  LiF 1A channel  by a factor of $\sim 2$. In the  LiF 2B channel observation, this feature is measured to have a strength of  $W_{observed} = 172.2~{\pm}~23.6$~m{\AA} over the wavelength interval of 1014.8 {\AA} - 1015.2 {\AA}, whereas in the LiF 1A channnel  the equivalent width obtained from integrating over the same interval is  $W_{observed} = 13.0 ~{\pm}~15.0$~m{\AA}. Since the feature is virtually absent in the higher $S/N$ LiF 1A channel observation, we classify this as a spurious feature. 

The line parameters, column density and Doppler width, for each of the above detected lines observed by FUSE were obtained using the apparent optical depth method of \citet{savage91} with $f$-values taken from \citet{verner96}. These measurements are listed in Table 1. The velocity interval over which the integrations of the optical depths were carried out are marked in Figure 1. In the case of {\OII}, the measurement is an upper limit on the column density, derived from the 3~$\sigma$ equivalent width non-detection, assuming that the line is on the linear part of the curve of growth. The 1~$\sigma$ uncertainty in the apparent column density also combines photon counting, continuum placement and fixed pattern noise errors in quadrature. Single component Voigt profiles were also fit to each of the detected lines and the measurements are tabulated in Table 2. The profile fitting procedure takes only the photon counting error into account. The significance of the {\NeVIII} column density thus measured is 5.8~$\sigma$, which is larger than the significance measured from the line equivalent width with conservative estimates of the error. The two {\CIV} lines detected in the FOS observations yield a doublet ratio of $2.00~{\pm}~0.45$. Thus line saturation is not likely. A curve of growth analysis yields log~$[{\CIV}~{\cmsq}] = 13.87^{+0.07}_{-0.04}$ and $b > 40$~{\kms}. The inferred value of $b$ is consistent with the {\OIV} and {\NIV} line widths. 

There is a strong {\Lya} feature ($W_r = 0.71~{\pm}~0.08$~{\AA}, Bahcall {\etal}1993) detected in the G190H FOS spectrum of this quasar at $z=0.4541$  for which the associated {\OIII}~$\lambda 702$~{\AA}, if present, would be at the same wavelength as the {\NeVIII}~$\lambda 770$~{\AA} line. However, for this {\Lya} absorber no associated metal lines are seen, including those of {\OIV}~$\lambda 787.71$~{\AA}, and {\NIV}~$\lambda 765.15$~{\AA},  which are normally expected. Hence we rule this out as a source of contamination for the {\NeVIII} feature. We also note that the wavelength region containing the {\NeVIII}~$\lambda 770$~{\AA} line (redshifted to $1021.3$~{\AA}) is not critically affected by ISM atomic lines or contamination from H$_2$ absorption.

\section{Origin of the Ionization in the Multiphase System}

In order to assess the physical conditions in the absorber, and the dominant ionization mechanisms, we first considered the possibility of the gas being photoionized by an extragalactic ionizing radiation field as modeled by \citet{hm96}. With the measured values of column densities as constraints, we constructed photoionization models using the ionization code Cloudy [ver 08.00$\beta$, \citet{ferland98}] for a range of values for ionization parameter (log~$U$), metallicity ([Z/H]), and {\HI} column density.  In the models we assume a solar abundance pattern of elements. We adopt solar O, N and Ne abundances from \citet{holweger01} and \citet{asplund04}. Figure 2 shows the Cloudy photoionization curves displaying the model column densities for the various ionization species under consideration. Our models predict that the intermediate ionization species {\OIII}, {\OIV}, {\NIV} and {\CIV} are consistent with having an origin in a single photoionized phase. The ionization parameter and gas density in this phase are constrained by $N(\OIII)/N(\OIV) \sim 1$.  As is evident from Figure 2, such a column density ratio is predicted for log~$U = -2.0$, which corresponds to $n_{\H} = 10^{-4}$~{\cc} for the given intensity of the incident ionizing radiation. 

The absence of information on {\HI} restricts our ability to reliably determine the abundances in the absorber. For the above estimated density, there is a range of $N(\HI)$ values for which a certain combination with metallicity can produce models that are consistent with the constraints set by the intermediate ions. A column density of $N(\HI) = 10^{15}$~{\cmsq}, and [Z/H] = -0.38 yields one such acceptable single phase solution for the photoionized medium (see Figure 2). With a decrease in $N(\HI)$, the metallicity has to increase to recover the observed metal line strengths. This sets a limit on the {\HI} column density for the photoionized phase. At low values, such as $N(\HI)=10^{14.6}$~{\cmsq}, the best fit model will have a [Z/H]$ > 0$, which is unusual for intergalactic gas. However, we note that  absorption systems with super-solar metallicities are known to exist over a range of {\HI} column densities \citep{charlton03,simcoe04,aracil06,prochaska06,misawa08}. In the same way, an upper limit for $N(\HI)$ can be  established by limiting the path length of the absorber to physically realistic values. At $N(\HI) > 10^{16}$~{\cmsq}, a density of $n_{\H} = 10^{-4}$~{\cc} would imply a large line of sight thickness of $L > 150$~kpc and a temperature of $T > 30,000$~K which is inconsistent with photoionization. Thus, for the photoionized phase of the absorber, the {\HI} is constrained to within $10^{14.6} < N(\HI) < 10^{16}$~{\cmsq} with a corresponding large range for metallicity of $0.1 > $~[Z/H]~$ > -4.0$. One of the possible single phase solutions shown in  Figure 2 is defined by a [Z/H] = -0.38, $n_{\H} = 1.3 \times 10^{-4}$~{\cc}, $N(\H) = 2 \times 10^{18}$~{\cmsq}, $T = 1.67 \times 10^4$~K, $f_{\HI} = 4.7 \times 10^{-4}$ and a path length of $L = 5.6$~kpc. This model yields a low chi-square statistic between the observed value of column density and those predicted by the model. Similar acceptable models can be derived for other combinations of $N(\HI)$ and metallicity within the range of values given above. Finally, we note that the data does not suggest any significant variation in the relative elemental abundance pattern from solar. The single phase model predicts that the N/O or C/O is within $\sim 0.2$~dex of solar abundances. 

A significant result from the above analysis is that the {\NeVIII} is not produced in the same temperature and density medium as the {\OIII} and {\OIV}. For the best fit single phase photoionization model discussed above, the derived log~$N(\NeVIII)$ is $\sim 4$~dex lower than the observed value. Assuming the {\NeVIII} phase to be of similar metallicity as the low-density photoionized gas (a valid assumption, since they are part of the same system), in order to recover the observed $N(\NeVIII)$ through photoionization,  a gas phase medium with $N(\H) \sim 10^{20}$~{\cmsq} and a very low density of $n_{\H} \sim 5.0 \times 10^{-6}$~{\cc} is required. This implies a line of sight thickness of $L \sim 5.6$~Mpc, an exceedingly large value for a single absorber. Moreover, the line width due to Hubble broadening from such a wide path length will be a factor of $\sim 8$ larger than the measured Doppler width of $49.8~{\pm}~5.5$~{\kms} for the {\NeVIII} line (see Table 1). The models show that for the $N({\H})$  to be lowered by even $\sim 1$~dex from the above value, and thus reduce the path length, the metallicity has to be supersolar ([Z/H] $> 1$). 

The most tenable idea for the origin of {\NeVIII} is collisional ionization in hot gas ($T > 5 \times 10^5$~K). Under collisional ionization equilibrium (CIE), the ionization fraction of {\NeVIII} depends entirely on the temperature, and peaks at T $= 7 \times 10^{5}$~K \citep{sutherland93}. Within the temperature regime of $(0.4 -  6) \times 10^{5}$~K, {\NeVIII} undergoes a steep increase of $\sim 6$~dex in ionic density. The ion is therefore a very sensitive probe of temperature in a collisionally ionized medium. For our detected absorber, we do not have sufficient constraints to determine the exact physical conditions in its collisionally ionized phase. Either direct information on the broad {\HI} component associated with this phase, or observation of another high ion (such as {\OVI}) is necessary so that the observed and expected values under collisional ionization equilibrium can be compared for a specific temperature in the gas. Yet, as we illustrate in the next section, some insights can be gained  through comparison with $z=0.20701$ {\NeVIII} system described in \citet{savage05}. 

\section{Similarity Between The Two Systems with Detected {\NeVIII} Absorption}

The $z=0.20701$ system along the HE~$0226-4110$ sight line was identified as a multi-phase photoionized and collisionally ionized absorber by Savage {\etal}(2005). Ionization models with strong constraints from several line measurements ({\Lya} to {\Lyth}, {\CIII}, {\OIII}, {\OIV}, {\OVI}, {\NIII}, {\SiIII} and {\SVI}) showed that the absorber is tracing a multiphase medium composed of a relatively cool (T~$\sim 2 \times 10^4$~K) photoionized phase of modest ionization detected via the intermediate ionization species, and a warm (T~$\sim 5 \times 10^5$~K) collisionally ionized phase traced via {\OVI} and (more importantly) {\NeVIII}. The {\OIV} column density in \citet{savage05} is given as a lower limit ($N_a > 3.2 \times 10^{14}$~{\cmsq}) due to the  {\OIV}~$\lambda 788$~{\AA} line blending with ISM {\OI}~$\lambda 950$~{\AA}. The measurement for {\NIV} is omitted due to the redshifted $\lambda 765$~{\AA} line being placed in the Galactic high velocity cloud portion of the {\HI}~$\lambda 923$~{\AA} feature. The {\OIII} column densities however are firm measurements in both systems and are within 1~$\sigma$ of each other (see Table 1). The density estimated for the photoionized phase in both absorbers are  comparable (n$_{\H} \sim 10^{-4}$~{\cc}), and the ionization correction relatively large ($f_{\HI} \sim 10^{-4}$). 

It is interesting to note that the {\NeVIII} in the two systems are similar in observed properties. The $N(\NeVIII)$ in the two absorbers are within 0.13 dex of each other, and the {\NeVIII}~$\lambda 770$~{\AA} rest-frame equivalent width within $1.5 \sigma$ (see Table 1). In \citet{savage05}, a collisionally ionized phase was found necessary to reproduce the observed {\NeVIII} and {\OVI}. The additional constraint from {\OVI} [$N(\NeVIII)/N(\OVI) = 0.3$] was used to derive the temperature in the warm plasma ($T = 5.4 \times 10^5$~K). As it turns out, {\OVI} is not as sensitive a probe of shock-heated gas as {\NeVIII}. The {\OVI} can be produced in both low density cool photoionized medium as well as warm collisionally ionized gas \citep[e.g.][]{savage02,danforth05,tripp08}. 

For the photoionized phase in the system that we present, the model predicts $N(\OVI) = 1.6 \times 10^{12}$~{\cmsq} (see Figure 2). Such a low column density feature can only  be detected at sufficiently high S/N in the FUV. In \citet{savage05}, photoionization significantly under-predicted the amount of {\OVI} required to explain the data for the lower states of ionization. The amount of {\OVI} from the collisionally ionized gas was $\sim 2$ dex larger compared to the contribution from the photoionized medium. It is therefore important to observe the redshifted {\OVIdblt} lines associated with the $z=0.32566$ system,  to determine with precision the conditions in the {\NeVIII} gas. The {\HI} associated with such hot gas is going to be broad ($b \sim 100$~{\kms}), and also shallow ($\tau = 0.024$, if $N$(H)$=10^{20}$~{\cmsq}) since the gas is also heavily ionized with $f_{\HI} \sim 10^{-6}$ \citep{savage05}.  Direct {\HI} observations will potentially confirm the presence of a such a broad component consistent with a collisional ionization origin. We note that 3C~263 is slated for observation with the {\it Cosmic Origins Spectrograph} (COS).

In {\OVI} intervening absorbers with solar elemental abundance ratios, collisional ionization equilibrium conditions can give rise to detectable amounts of {\NeVIII} [$N(\NeVIII) \gtrsim 10^{14}$~{\cmsq}] at temperatures T $> 5 \times 10^5$~K \citep{sutherland93,savage05}. In \citet{savage05}, the $W_r(\OVI~\lambda 1032)/W_r(\NeVIII~\lambda 770) \sim 1$ was recovered from a collisionally ionized gas phase with T $ = 5.4 \times 10^5$~K. As explained earlier, separate ionization mechanisms are capable of producing {\OVI}, which implies that {\NeVIII} might not always be associated with {\OVI} systems. For the four {\OVI} absorbers in \citet{richter04} that had coverage, {\NeVIII} was reported as a non-detection at the 3~$\sigma$ significance. In Lehner {\etal}(2006), only one out of four {\OVI} absorbers had a $\geq 3$~$\sigma$ significance {\NeVIII} feature \citep{savage05}. In the low-z universe ($z < 0.5$), the rate of incidence of {\OVI} systems with $W_r > 30$~m{\AA} per unit redshift is estimated to be $dN(\OVI)/dz \sim 15$ \citep[e.g.][]{lehner06,tripp08}. Based on the above (limited) information on the association of {\NeVIII} with {\OVI}, we estimate the number of intervening {\OVI} absorbers ($W_r > 30$~m{\AA}) with {\NeVIII} to be $dN/dz \sim 1/7 \times dN(\OVI)/dz \sim 2.1$ for $z < 0.5$\footnote{Out of the eight systems in \citet{richter04} and \citet{lehner06} combined, one of the {\OVI} systems in \citet{richter04} has $W_r(\OVI~\lambda 1032) < 30$~m{\AA} and thus is excluded from the estimation. Hence the scaling factor for dN/dz is one-seventh.}. The {\NeVIII} detection reported in this paper has been excluded from the above calculation since the FUSE spectral window on 3C~263 was inadequate for a simultaneous search for {\OVI} and associated {\NeVIII} over any path length. 

Although the estimated value of $dN/dz \sim 2.1$ for {\NeVIII} absorbers is highly uncertain, it is still interesting to use this number to estimate the possible baryonic content in these absorbers.  Using equation $9$ of \citet{rao00} and the assumption that the typical {\NeVIII} system has the properties of the system at $z=0.20701$ seen toward HE~$0226-4110$ with $T = 5.4 \times 10^5$~K, $N(\H) = 8.3 \times 10^{19}$~{\cmsq}, and [Ne/H]$=-0.5$ \citep{savage05}, we evaluate the baryonic contribution to the closure density by {\NeVIII} systems to be $\Omega_b (\NeVIII) \sim 0.0026$\footnote {For $H_o = 70$~{\kms}~Mpc$^{-1}$}. With $\Omega_b$(total) $= 0.045$, and $\Omega_b$(galaxies) $= 0.0035$ \citep{fukugita04}, the {\NeVIII} systems may thus contain $\sim 6$~\% of the baryons. This value is comparable to the baryonic content of galaxies at low-$z$. Note that the above estimate depends on the frequency of incidence of these absorbers, the assumed metallicity, and the correctness of the derived $N(\H)$ for the collisionally ionized gas. It is independent of where the {\NeVIII} absorbers are actually located. The result illustrates the importance of obtaining detailed information on the properties of {\NeVIII} systems and their frequency of occurrence in the universe. 

\section{Conclusion}

We have presented the detection of {\NeVIII}~$\lambda 770$~{\AA} line at $z=0.32566$ in the FUSE spectrum of 3C~$263$. The measurement is significant at the $3.9 \sigma$ level, making it one of only two $\geq 3 \sigma$ detections of this ion in intervening absorbers. We also detect {\OIII}, {\OIV} and {\NIV} associated with this system in the FUSE spectrum and {\CIVdblt} lines in the FOS data. These intermediate ionization species are consistent with a single phase photoionization solution in a medium with a low density ($n_{\H} \sim 10^{-4}$~{\cc}). The ratio between the measured column densities of {\CIV}, {\NIV}, and {\OIV} compared with the photoionization model predictions indicate that the relative abundances of C, N, O in this phase are likely within $\sim 0.2$~dex of solar abundances. The {\NeVIII} in this absorber is inconsistent with having an origin in a medium that is predominantly photoionized.  The absorber is a multi-phase system in which the {\NeVIII} is produced via collisional ionization in a warm plasma, whose temperature corresponds to $T \sim (0.5 - 1) \times 10^6$~K. More detailed constraints on the physical state of this gas can be obtained only through direct observations of the {\OVI} and {\HI} absorption associated with this system. 

To better understand the origin of this absorber, it is also important to determine its actual physical location.  Measures of the redshift of galaxies in the general direction of 3C~$263$ would allow the determination of whether the absorber is tracing the extended halo of an intervening galaxy, intragroup gas, or possibly a structure of the WHIM connecting galaxies. The analysis presented here, in general, confirms the importance of {\NeVIII} as a new probe of collisionally ionized gas in the low-$z$ universe. The prospects for discovering other such systems is going to be considerably augmented in the HST/$COS$ era.

\vspace{0.5in}

We appreciate the substantial efforts required by the {\it FUSE} Operations Team to extend the life of {\it FUSE} and to obtain the long 196 ksec integration on 3C 263.   This research has been supported by the NASA {\it Cosmic Origins Spectrograph} Program through a sub-contract to the University of Wisconsin, Madison  from the University of Colorado, Boulder.  We also thank an anonymous referee for several useful suggestions towards improving the scope of this paper.  B.P.W acknowledges support from NASA grant NNX-07AH426.


\begin{deluxetable}{lcccc}
\tabletypesize{\footnotesize}
\tablecaption{\textsc{Metal Line Measurements}}
\tablehead{
\colhead{ion} &
\colhead{$\lambda_{rest}$~({\AA})} &
\colhead{$W_r$~(m{\AA})} &
\colhead{log$~N$~({\cmsq})} &
\colhead{$b$~({\kms})}
}
\startdata
\cutinhead{3C263, $z=0.32566$ system}
{\NeVIII} & $770.409$ & $47.0~{\pm}~11.9$ [{\scriptsize 9.6, 4.9, 5.0}]$^a$ & $13.98^{+0.10}_{-0.13}$ & $49.8~{\pm}~5.5$\\	
{\OIII} & $832.927$ & $95.1~{\pm}~10.1$ [{\scriptsize 7.9, 3.8, 5.0}]$^a$ & $14.25^{+0.05}_{-0.06}$ & $42.0~{\pm}~2.6$ \\
{\OIV} & $787.711$ & $79.2~{\pm}~8.8$ [{\scriptsize  6.4, 3.5, 5.0}]$^a$ & $14.21^{+0.05}_{-0.06}$ & $35.8~{\pm}~1.9$ \\
{\NIV} &  $765.148$ & $97.5~{\pm}~14.6$ [{\scriptsize 12.0, 6.6, 5.0}]$^a$  & $13.57^{+0.07}_{-0.08}$ & $51.7~{\pm}~4.4$ \\
{\OII} & $834.465$ & $< 29.8^b$ & $< 13.2^b$ & $ -- $  \\
{\CIV}$^{ c}$ & $1548.204$ & $273.3~{\pm}~28.5$[{\scriptsize 24.1, 15.2, 0}] & \multirow{2}{*}{$13.87^{+0.07}_{-0.04}$} & \multirow{2}{*}{$> 40$} \\
{\CIV}$^{c}$ & $1550.781$ & $136.4~{\pm}~27.8$[{\scriptsize 24.9, 12.4, 0}]  \\
\cutinhead{HE~$0226-4110$, $z=0.20701$ system$^d$} \\
{\NeVIII} & $770.409$ & $32.9~{\pm}~10.5$ & $13.85^{+0.12}_{-0.17}$ & $28.3~{\pm}~8.8$ \\
{\OIII} & $832.927$ & $90.8~{\pm}~12.4$ & $14.30^{+0.06}_{-0.06}$ & $23.1~{\pm}~3.1$ \\
{\OIV} & $787.711$ & $> 91.9$ & $> 14.50$ & $> 17.4$ \\
{\OVI} & $1031.926$ & $169~{\pm}~15.3$ & $14.36~{\pm}~0.05$ & $29~{\pm}~2.1$ \\
{\OVI} & $1037.617$ & $112~{\pm}~10.4$ & $14.38~{\pm}~0.04$ & $31.3~{\pm}~1.7$ 
\enddata
\label{tab:tab1}
\tablecomments{\scriptsize The apparent column density method of Savage \& Sembach (1991) was used to derive the column density ($N_a$), except for {\CIV} and {\OII} where the values are based on curve of growth method. The integration intervals are marked in the Figure 1 system plot. The {\NeVIII}~$\lambda 780$~{\AA} line is affected by strong contamination from a Lyman-$\gamma$ feature at $z = 0.4541$. The 1~$\sigma$ error given in these measurements is a combination in quadrature of the statistical error, continuum placement error and fixed pattern noise.}
\tablenotetext{a}{\scriptsize The values in the square brackets are the statistical uncertainty, continuum placement error, and fixed-pattern noise for the respective measurements.}
\tablenotetext{b}{\scriptsize This is a non-detection and the values are therefore $3~\sigma$ upper limits.}
\tablenotetext{c}{\scriptsize This line is detected in the FOS G190H grating spectrum of 3C~263 \citep{bahcall93}. We derive the equivalent widths by integrating over the velocity interval [-250, 250]~{\kms}. The column density was derived using curve of growth analysis.}
\tablenotetext{d}{\scriptsize These measurements are taken from \citet{savage05}. The {\HI} column density measurement from the best-fit profile for this system has two components at $v = 5$~{\kms} and $v = -24$~{\kms}, with log [$N~{\cmsq}$] $= 14.89~{\pm}~0.04$ and $15.06~{\pm}~0.03$ respectively and corresponding $b = 35.9~{\pm}~1.1$~{\kms} and $17.4 ~{\pm}~1.0$~{\kms} respectively. These measurements are limited by the lack of information on the true component structure seen for the Lyman series lines. The numerous other ions detected for this system are not listed here.}
\end{deluxetable}

\begin{deluxetable}{lcccc}
\tablecaption{\textsc{Voigt Profit Fit Results}}
\tablehead{
\colhead{ion} &
\colhead{$v$~({\kms})} &
\colhead{$\chi_{\nu}^2$} &
\colhead{log~$N$~({\cmsq})} &
\colhead{$b$~({\kms})}
}
\startdata
{\NeVIII} & $-13.82$ & $0.842$ & $14.06~{\pm}~0.08$ & $52.4~{\pm}~8.8$ \\
{\OIII} & $4.90$ & $0.809$ & $14.24~{\pm}~0.03$ & $41.6~{\pm}~3.4$ \\
{\OIV} & $0.54$ & $0.424$ & $14.26~{\pm}~0.02$ & $35.0~{\pm}~2.1$ \\
\enddata
\label{tab:tab2}
\tablecomments{Single component Voigt profile parameterization of the lines. The 0~{\kms} corresponds to the system redshift given by the {\OIV} line profile. The fitting of the line profiles was carried using the routines AUTOVP \citep{dave97} and MINFIT \citep{cwc03}. The 1~$\sigma$ uncertainty in these derived quantities account only for the statistical (photon counting) error. Therefore the column density errors listed here are substantially smaller than the more realistic errors given in Table 1.}
\end{deluxetable}


\begin{figure*}
\vspace{1in}
\epsscale{0.85}
\begin{center}
\plotone{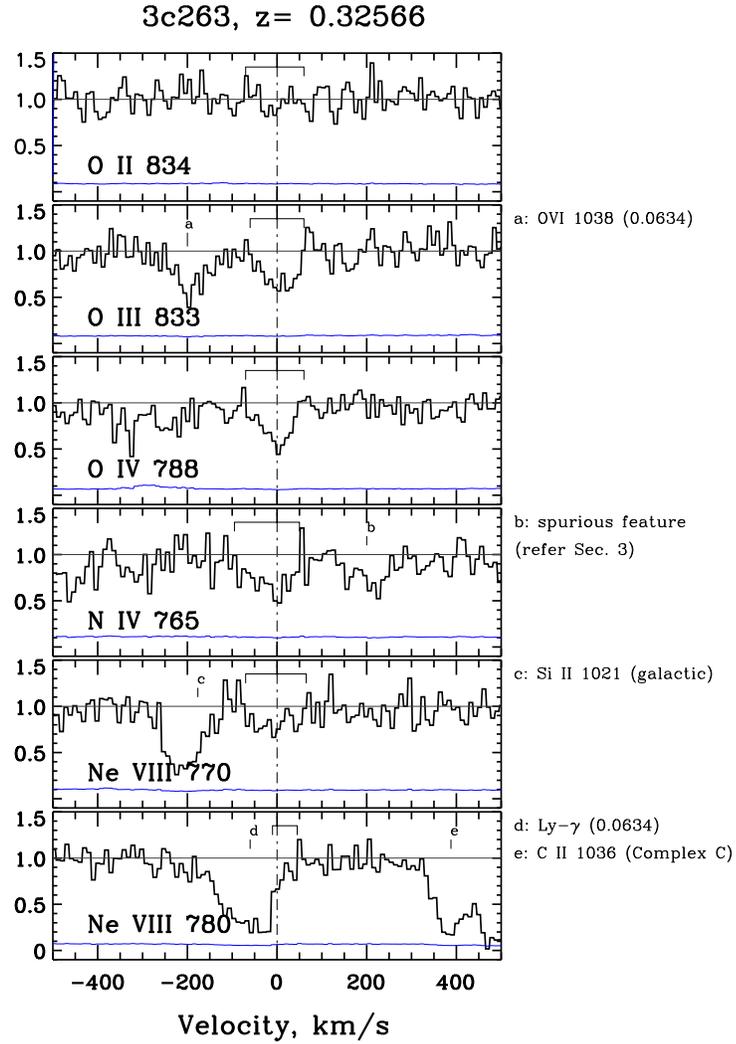}
\end{center}
\vspace{0.5in}
\protect
\caption{\small System plot of the $z=0.32566$ absorber in the FUSE 3C~$263$ sight line. The vertical axis is the continuum normalized flux, the horizontal axis the rest-frame velocity of the lines detected, with $0$~{\kms} corresponding to the redshift of the system given by the {\OIV} line. The 1~$\sigma$ photon counting error spectrum per bin (see Sec 2) is plotted in the bottom of each panel. Above each feature is marked the velocity interval for the apparent column density integrations. The {\OII}~$\lambda 834$~{\AA} line is not detected at $\geq 3$~$\sigma$ significance, and the equivalent width limit is derived by integrating over the same velocity window as {\OIV}. The {\NeVIII}~$\lambda 780$~{\AA} is heavily blended with the {\HI}~$\lambda 972$~{\AA} of an absorber at $z=0.0634$. The few pixels on the red region of this feature appears to be unblended from comparing by-eye with the {\NeVIII}~$\lambda 770$~{\AA}, and we have marked this in the spectrum for convenience. The other dominant absorption features in the various plot windows are also identified and labeled. In the {\NeVIII}~$\lambda 780$~{\AA} panel, the strong features at $548$~{\kms} and $388$~{\kms} are the {\CII}~$\lambda 1036$~{\AA} absorption from the Galaxy at $v = 0$~{\kms} and high velocity cloud Complex C at $v=-160$~{\kms}.}
\label{fig:1}
\end{figure*}

\begin{figure*}
\vspace{1in}
\epsscale{1.0}
\begin{center}
\plotone{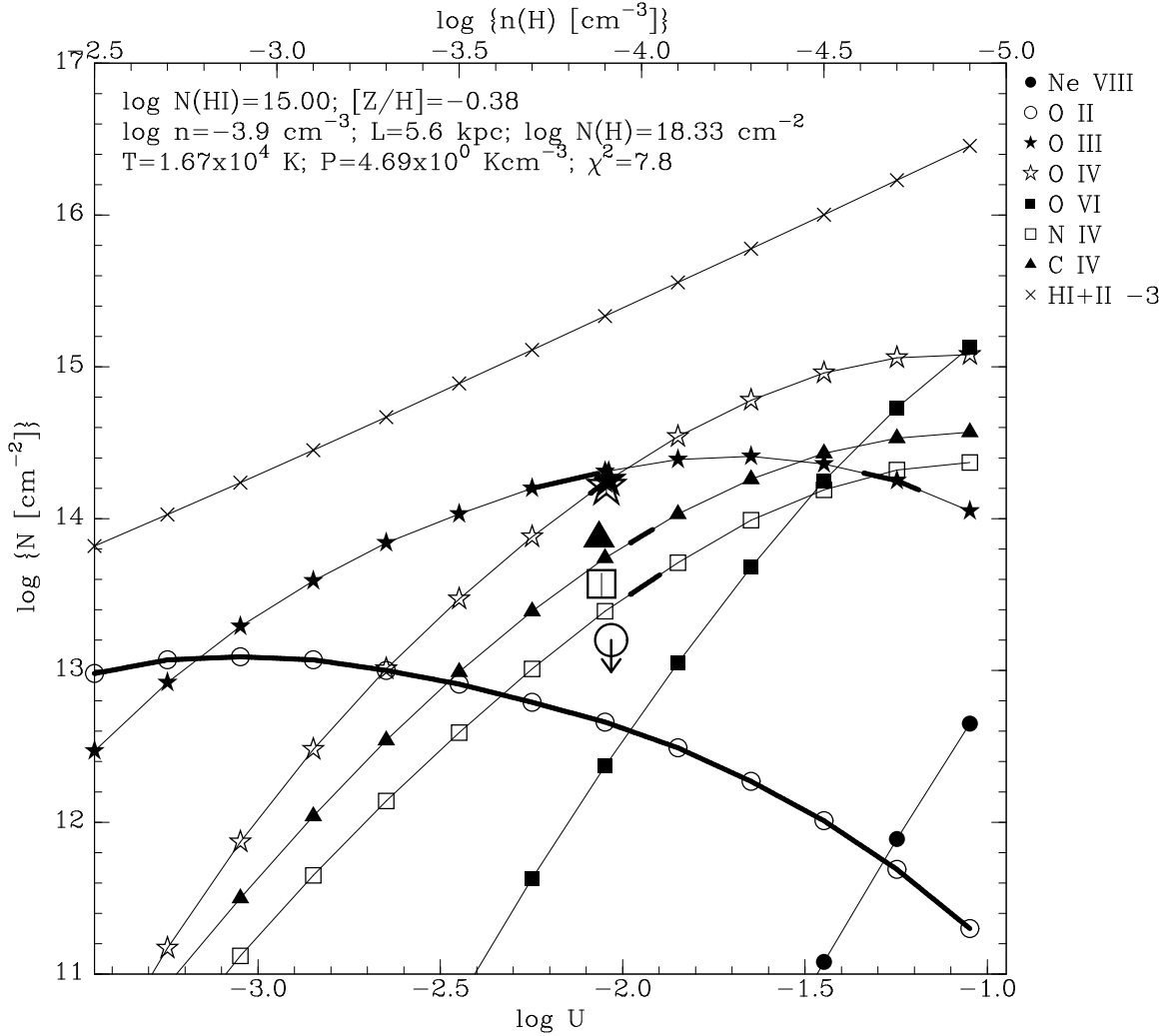}
\end{center}
\protect
\caption{\small A photoionization model for the $z=0.32566$ system. The curves show the run of column density with ionization parameter (log~$U$) for a given $N(\HI)$ and [Z/H]. The predictions of this relatively best-fit model ($\chi^2 = 7.8$, fitted to 4 ions) are given within the figure panel. The top horizontal axis is the density of the photoionized medium derived using the expression log~$n_{\H} =$~log~$n_{\gamma} -$~log~$U$, where $n_{\gamma}$ is the number density of ionizing photons with $h\nu \geq 13.6$~eV from the background radiation field at $z = 0.32566$ modeled by Haardt \& Madau (1996). The measured column densities (see Table 1) are marked using big symbols. The {\OII}, {\OIII}, {\OIV}, and {\NIV} constraints are from FUSE and {\CIV} from FOS data. The thick lines on the curves are the 1~$\sigma$ uncertainty in the derived column density of each ion. The total $N(\H) = N(\HI) + N(\HII)$ is scaled down by 3 dex to fit within the plot window. The measured {\OII} column density is an upper limit and is therefore marked with a downward pointing arrow. The {\NeVIII} column density is not marked in the figure as the photoionization models are unable to recover it from this same phase. For comparison, the model column densities for {\OVI} in this photoionized phase are also shown. The amount of {\OVI} expected in the photoionized gas is only $N(\OVI) \sim 10^{12.2}$~{\cmsq}.}
\label{fig:2}
\end{figure*}
\clearpage

\end{document}